\documentclass[letterpaper, 10 pt, conference]{ieeeconf}

\usepackage[utf8]{inputenc}
\usepackage[T1]{fontenc}
\usepackage[english]{babel}

\usepackage{mathtools}
\usepackage{amssymb} 

\usepackage{bm} 

\usepackage{nccmath}

\usepackage[version=3]{mhchem} 
\usepackage{esint} 
\usepackage{dsfont} 
\usepackage[makeroom]{cancel} 
\usepackage{empheq}
\usepackage{physics}
\usepackage[binary-units=true,per-mode=symbol]{siunitx}

\usepackage{graphicx}
\usepackage{float}

\usepackage{subcaption}

\usepackage[export]{adjustbox} 

\usepackage{booktabs}
\usepackage{makecell}
\usepackage{threeparttable}
\usepackage{multirow}
 
\usepackage[dvipsnames]{xcolor}

\usepackage[nameinlink,capitalise,noabbrev]{cleveref}
\crefformat{equation}{(#2#1#3)}
\Crefformat{equation}{Equation (#2#1#3)}
\crefrangeformat{equation}{(#3#1#4) to~(#5#2#6)}
\crefmultiformat{equation}{(#2#1#3)}%
{ and~(#2#1#3)}{, (#2#1#3)}{ and~(#2#1#3)}
\Crefmultiformat{equation}{Equations (#2#1#3)}%
{ and~(#2#1#3)}{, (#2#1#3)}{ and~(#2#1#3)}

\usepackage[noadjust]{cite} 

\usepackage{pgf,tikz,pgfplots}
\pgfplotsset{compat=1.10}
\usetikzlibrary{shapes,arrows,calc,arrows.meta,
patterns,backgrounds,positioning,fit,chains}
\tikzset{every picture/.style=black}
\usepackage[american]{circuitikz}

\usepackage{relsize}

\tikzset{
    declare function={
        atan3(\a,\b)=ifthenelse(atan2(0,1)==90, atan2(\a,\b), atan2(\b,\a));
    },
    kinky cross radius/.initial=+.125cm,
    @kinky cross/.initial=+,
    kinky crosses/.is choice,
    kinky crosses/left/.style={@kinky cross=-},
    kinky crosses/right/.style={@kinky cross=+},
    kinky cross/.style args={(#1)--(#2)}{
        to path={
            let \p{@kc@}=($(\tikztotarget)-(\tikztostart)$),
            \n{@kc@}={atan3(\p{@kc@})+180} in
            -- ($(intersection of \tikztostart--{\tikztotarget} and #1--#2)!%
            \pgfkeysvalueof{/tikz/kinky cross radius}!(\tikztostart)$)
            arc [ radius     =\pgfkeysvalueof{/tikz/kinky cross radius},
            start angle=\n{@kc@},
            delta angle=\pgfkeysvalueof{/tikz/@kinky cross}180 ]
            -- (\tikztotarget)
        }
    }
}

\tikzset{three sided/.style={
        draw=none,
        append after command={
            [shorten <= -0.5\pgflinewidth]
            ([shift={(-1.5\pgflinewidth,-0.5\pgflinewidth)}]				  \tikzlastnode.north east)
        edge([shift={( 0.5\pgflinewidth,-0.5\pgflinewidth)}]				  \tikzlastnode.north west) 
            ([shift={( 0.5\pgflinewidth,-0.5\pgflinewidth)}]				  \tikzlastnode.north west)
        edge([shift={( 0.5\pgflinewidth,+0.5\pgflinewidth)}]				  \tikzlastnode.south west)            
            ([shift={( 0.5\pgflinewidth,+0.5\pgflinewidth)}]				  \tikzlastnode.south west)
        edge([shift={(-1.0\pgflinewidth,+0.5\pgflinewidth)}]				  \tikzlastnode.south east)
        }
    }
}

\usepackage{psfrag}
\usepackage{pstool} 

\makeatletter
\newcommand{\getfontsize}{\f@size pt}
\makeatother

\DeclareMathOperator{\E}{\mathbb{E}}
\DeclareMathOperator{\Var}{Var}

\newcommand*\diff{\mathop{}\!\mathrm{d}}

\newcommand*\kB{k}

\newcommand*\VDD{V_{\mathrm{DD}}}

\newcommand*\iDD{i_{\mathrm{DD}}}

\newcommand*\fmax{f_{\mathrm{max}}}

\newcommand{\dt}{\diff{t}}

\newcommand*{\x}{x}

\newcommand*\iex{i_{\mathrm{ex}}}

\newcommand*\vM{v_{\mathrm{m}}}
\newcommand*\vGNA{v_{\mathrm{GNA}}}
\newcommand*\vGK{v_{\mathrm{GK}}}

\newcommand*\CM{C_{\mathrm{m}}}
\newcommand*\CK{C_{\mathrm{K}}}

\definecolor{k}{rgb}{0 0 0}
\definecolor{r}{rgb}{1 0 0}
\definecolor{g}{rgb}{0 1 0}
\definecolor{b}{rgb}{0 0 1}
\definecolor{orange}{rgb}{1,0.7,0}
\definecolor{darkOrange}{RGB}{255 120 0}
\definecolor{c}{rgb}{0 1 1}
\definecolor{cc}{RGB}{64 224 208}
\definecolor{m}{rgb}{1 0 1}
\definecolor{khaki}{RGB}{128 128 0}
\definecolor{deepskyblue}{RGB}{0 191 255}
\definecolor{darkMagenta}{rgb}{0.5 0 0.5}
\definecolor{chocolateBrown}{RGB}{98 52 18}
\definecolor{lightBrown}{RGB}{189 154 122}
\definecolor{mybrown}{RGB}{127 37 0}
\definecolor{bordeaux}{RGB}{131 41 85}
\definecolor{violet}{RGB}{127 0 255}
\definecolor{myGreen}{RGB}{134,180,44}
\definecolor{gray_gate}{RGB}{211,208,205}
\definecolor{yellow_oxide}{RGB}{244,231,164}
\definecolor{color_mix}{rgb}{0.7510 0.2510 0.2510}

\definecolor{h}{rgb}{0 0 0}

\definecolor{l}{rgb}{0 0 1}

\colorlet{Iexcolor}{r}
\colorlet{VMcolor}{deepskyblue}
\colorlet{VGNAcolor}{darkMagenta}
\colorlet{VGKcolor}{b}

\definecolor{h}{rgb}{0 0 0}
\definecolor{l}{rgb}{0 0 0}

\graphicspath{{figures/}}

\makeatletter
\def\maketag@@@#1{\hbox{\m@th\normalfont\normalsize#1}}
\makeatother

\IEEEoverridecommandlockouts 

\title{\LARGE \bf
Interplay between Excitability and Noise in
Analog Spiking Neurons
}

\author{Léopold Van Brandt, Alon Ascoli, Michele Bonnin, Grégoire Brandsteert,\\ Denis Flandre and Jean-Charles Delvenne
\thanks{
Léopold Van Brandt, Grégoire Brandsteert, Denis Flandre and Jean-Charles Delvenne are with the ICTEAM Institute, UCLouvain, Louvain-la-Neuve, Belgium
{\tt\small leopold.vanbrandt@uclouvain.be}} %
\thanks{
Alon Ascoli and Michele Bonnin are with the Department of Electronics and Telecommunications, Politecnico di Torino, Turin, Italy}
\thanks{
Jean-Charles Delvenne is also a FY2025 JSPS fellow and Invited Professor at the Department of Physics, Kyoto University, Kyoto, Japan} 
\thanks{
This research was supported by the European Research Council (ERC) under the European Union’s Horizon AG research and innovation program (ERC-Synergy, SWIMS, 101119062), the Research Projects “Thermodynamics of Circuits for Computation” and “Stochastic Modelling of Present and Future Nonlinear Dynamical Electronic Devices and Circuits” of the F.R.S.-FNRS, and the SIDDARTA Concerted Research Action (ARC)
of the Fédération Wallonie-Bruxelles (Belgium).
} 
}

\begin{document}

\maketitle
\thispagestyle{empty}
\pagestyle{empty}

\begin{abstract} 


\textcolor{h}{
A spiking neuron is a dynamical system exhibiting a limit cycle (a spike) when subject to sufficiently excitatory input.
}
Famous neuroscience experiments by \textcolor{h}{Bryant and Segundo as well as} Mainen and Sejnowski revealed that the spike times of some biological neurons are more reliable when subject to time-varying stimuli compared to constant ones.
\textcolor{l}{
Provided with an industrial physics-based transient noise SPICE simulation framework compatible with foundry transistor compact models,} we have demonstrated similar behaviours for a CMOS analog spiking neuron designed for neuromorphic computing hardware platforms.
Under constant excitation current, the neuron operates in oscillatory regime and the 
\textcolor{h}{empirical periods (inter-spike intervals)}
suffer from accumulated \textcolor{h}{phase noise (jitter)} during successive cycles, mainly due to 
the thermal noise of transistors.
A time-varying input is reported to be capable of triggering 
output spikes 
according to a controlled state-transition mechanism, thereby significantly reducing the spike \textcolor{h}{time} jitter in the excitability regime.
\textcolor{h}{The physical limitation of the oscillatory or rate-coding regime in terms of reliability-dissipation tradeoff is further 
evidenced by a stochastic thermodynamic uncertainty relation, a theorem that applies to general overdamped dissipative stochastic dynamical systems.}

\end{abstract}

%
%
%


\section{Introduction}

\emph{Spike}-based neuromorphic computing \textcolor{h}{builds} on an information theory where the data are encoded on the timing of
\emph{events} rather than on amplitude of signals (currents and voltages)~\cite{Sepulchre2022_PIEEE}.
Major promises include energy efficiency 
and robustness against the non-idealities of the electronic devices, notably the intrinsic noise.
The spiking neurons are key circuit building blocks and can be implemented either in CMOS technology~\cite{Chicca2014_PIEEE,Querlioz2013,Sourikopoulos2017} or with volatile memristors~\cite{ascoli2025edge,zhao2025}.
The ultra-low-power (ULP) neuron architecture of \textcolor{l}{\cite{Sourikopoulos2017} (\cref{fig_neuron})} will be studied in this work since it is representative of more complex neurons like the adaptive exponential \textcolor{h}{(AdEx)} leaky integrate-and-fire cells~\cite{Chicca2014_PIEEE} integrated in state-of-the-art neuromorphic processors~\cite{Greatorex2025}.

{\color{h}

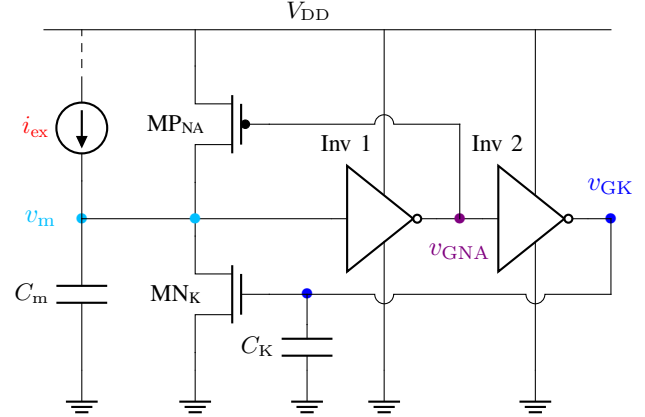
\begin{figure}[]
\centering
\begin{circuitikz}[american voltages, transform shape, line cap=rect, nodes={line cap=butt},scale=1]
\ctikzset{capacitors/scale=0.8}
\draw[VMcolor]
(0,0) node[label={[font=\normalsize,color=VMcolor]left:$\vM$}] (vM) {$\bullet$};
\newcommand\dyVDD{2.5}
\newcommand\dyGND{2.0}
\draw
(vM)++(0, +\dyVDD) node[] (VDDiex) {};
\draw
(vM)++(0, \dyVDD-0.5) node[] (belowVDDiex) {};
\draw
(vM)++(0,0.5) node[] (aboveiex) {};
\draw (vM)++(0,0.5) to[isource, invert, scale=0.8,font=\normalsize,label={\relscale{1.25}\color{Iexcolor}$\iex$}] (belowVDDiex) {};
\draw[black]
(vM.center) -- (aboveiex.center);
\draw[black,dashed]
(belowVDDiex.center) -- (VDDiex.center);
\draw
(vM)++(0, -\dyGND) node[ground,scale=1] (GNDiex) {};
\draw[black]
(vM.center) to[C,l_={\small$\CM$}] (GNDiex);
\draw
(vM) ++ (4,0) node[not port] (inv1) {};
\draw
(inv1) ++ (-0.5,+1) node[] (inv1label) {\small Inv 1};
\newcommand\dyanchorVDDinv{0.325}
\newcommand\dyanchorGNDinv{0.3}
\draw
(inv1)++(0, +\dyVDD) node[] (VDDinv1) {};
\draw[black]
(inv1)++(0,+\dyanchorVDDinv) -- (VDDinv1.center);
\draw
(inv1)++(0, -\dyGND) node[ground,scale=1] (GNDinv1) {};
\draw[black]
(inv1)++(0,-\dyanchorGNDinv) -- (GNDinv1.center);
\draw
(vM) ++ (6,0) node[not port] (inv2) {};
\draw
(inv2) ++ (-0.5,+1) node[] (inv2label) {\small Inv 2};
\draw
(inv2)++(0, +\dyVDD) node[] (VDDinv2) {};
\draw[black]
(inv2)++(0,+\dyanchorVDDinv) -- (VDDinv2.center);
\draw
(inv2)++(0, -\dyGND) node[ground,scale=1] (GNDinv2) {};
\draw[black]
(inv2)++(0,-\dyanchorGNDinv) -- (GNDinv2.center);
\draw[black]
(vM.center) -- (inv1.in);
\draw[black]
(inv1.out) -- (inv2.in);
\draw[VGNAcolor]
(vM) ++ (5,0) node[label={[font=\normalsize,color=VGNAcolor]below:$\vGNA$}] (vGNA) {$\bullet$};
 \draw[VGKcolor]
(vM) ++ (7,0) node[label={[font=\normalsize,color=VGKcolor]above:$\vGK$}] (vGK) {$\bullet$};
\draw[black]
(inv2.out) -- (vGK.center);
\newcommand\dxM{1.5}
\newcommand\dyMN{1.25}
\draw
(vM) ++ (\dxM,+\dyMN) node[pmos,rotate=180,yscale=-1] (MPNa) {};
\draw
(MPNa) ++ (-0.25,+0) node[] (MPNAlabel) {\small MP\textsubscript{NA}};
\draw
(vM) ++ (\dxM, +\dyVDD) node[] (VDDMPNa) {};
\draw[black]
(MPNa.S) -- (VDDMPNa.center);
\newcommand\dyMP{1.0}
\draw
(vM) ++ (\dxM,-\dyMP) node[nmos,rotate=180,yscale=-1] (MNK) {};
\draw
(MNK) ++ (-0.25,+0) node[] (MNKlabel) {\small MN\textsubscript{K}};
\draw
(vM) ++ (\dxM, -\dyGND) node[ground,scale=1] (GNDMNK) {};
\draw[black]
(MNK.S) -- (GNDMNK.center);
\draw[black]
(MPNa.D) -- (MNK.D);
\draw[VMcolor]
(vM) ++ (\dxM,0) node[] (vMM) {$\bullet$};
\draw[VGKcolor]
let \p1=(MNK.G) in (\x1,\y1) ++ (0.5,0) node[] (vGKK) {$\bullet$}
++ (0,-0.4) node[] (belowvGKK) {};
\draw[black]
(vGKK.center) -- (belowvGKK.center);
\draw
let \p1=(vGKK) in (\x1,-\dyGND) node[ground,scale=1] (GNDCK) {};
\draw[black]
(belowvGKK.center) to[C,l_={\small$\CK$}] (GNDCK);
\draw
 (vGNA |- MPNa.G) node[] (abovevGNA) {};
\draw[black]
(vGNA.center) -- (abovevGNA.center);
\draw[black]
(abovevGNA.center) to[kinky cross=(inv1)--(VDDinv1.center), kinky crosses=right] (MPNa.G);
\draw
 (vGK |- MNK.G) node[] (belowvGK) {};
 \draw
 (vGNA |- MNK.G) node[] (belowvGNA) {};
\draw[black]
(vGK.center) -- (belowvGK.center);
\draw[black]
(belowvGK.center) to[kinky cross=(inv2)--(GNDinv2.center), kinky crosses=left] (belowvGNA.center)
to[kinky cross=(inv1)--(GNDinv1.center), kinky crosses=left] (vGKK.center)
-- (MNK.G);
\draw
(vM) ++ (-0.5, +\dyVDD) node[] (VDDl) {}
(vM) ++ (+7, +\dyVDD) node[] (VDDr) {};
\draw[black]
(VDDl.center) -- (VDDr.center);
\draw[black]
(vM) ++ (3,\dyVDD+0.25) node[] (VDD) {\small$\VDD$};
\end{circuitikz}

\vspace{1mm}

\caption{
Diagram of an ultra-low-power CMOS analog spiking neuron
based on a \textcolor{h}{simplified} Morris-Lecar model~\cite{Sourikopoulos2017}.
\textcolor{l}{
The input is the excitatory current $\iex$, while the output is the node voltage $\vGK$.
}
}
\label{fig_neuron}
\end{figure}

}

In a recent paper~\cite{LASCAS2026}, we analysed the spiking mechanism and studied the excitability property of the neuron,
\textcolor{h}{ 
from the purely deterministic point of view,}
ignoring the intrinsic noise.
\textcolor{l}{
One goal of the present paper is to assess the effect of noise on the spike times, for different input signals, through \emph{transient noise simulations} (\texttt{NOISETRAN}) performed with an industrial SPICE tool~\cite{SSE2023,TCAS2025}.
The CMOS neuron design in $\SI{65}{\nano\meter}$ technology and all the deterministic parameters of the model of~\cite{Sourikopoulos2017} (\cref{fig_neuron}) are retained, especially the transistor dimensions and the supply voltage $\VDD = \SI{200}{\milli\volt}$.
By introducing the physical random noise intrinsic to the circuit,} we will 
\textcolor{l}{notably} 
demonstrate behaviours similar to biological neurons as reported \textcolor{h}{by~\cite{Bryant1976,Mainen1995}}.
The work~\cite{Kirby2021} replicated these neuroscience experiments for an artificial neuron made of discrete components on a circuit board but did not quantify the spike time uncertainty.

\textcolor{h}{
We report a few older works assessing the interplay between nonlinear neuronal dynamics and noise~\cite{Lansky1999,Shimokawa1999}.
They 
are theoretical in the sense that they rely on simplified mathematical models of the neuron, irrespective of the actual complexity of its physical implementations.
Moreover, in this framework, the noise is usually added as a white Gaussian process of fixed predefined intensity to the differential equations governing the neuronal dynamics.
This overlooks the state 
dependency of white noise sources in nonlinear electronic devices~\cite{Bonnin2025_TCAS} such as the MOS transistors in the circuit of \cref{fig_neuron}. 
In this regard, we believe that a more data-driven analysis, which would rely on industrial component models and circuit simulation tools, is
an important contribution, which is currently missing in the literature
}

The rest of this paper is structured as follows.
Given their major influence on this work, Mainen and Sejnowski~\cite{Mainen1995}'s experiments are revisited in \cref{section:Spike Timing in Biological Neurons: Neuroscience Experiments}.
Relying on rigorous transient noise simulations, we study the spike time reliability of the CMOS analog spiking neuron 
under different excitatory stimuli 
in \cref{section:Spike Timing in Analog Spiking Neurons: Numerical Investigation}.
We compare the results, acquired while exploring the neuron spiking behaviour in different regimes, in terms of timing variability in \cref{section:Results and Discussion}.
\Cref{section:Conclusions and Perspectives} draws the conclusions and opens perspectives for future work.

\section{Spike Timing in Biological Neurons: Neuroscience Experiments}
\label{section:Spike Timing in Biological Neurons: Neuroscience Experiments}

\begin{figure}
\centering
\includegraphics[width=\linewidth]{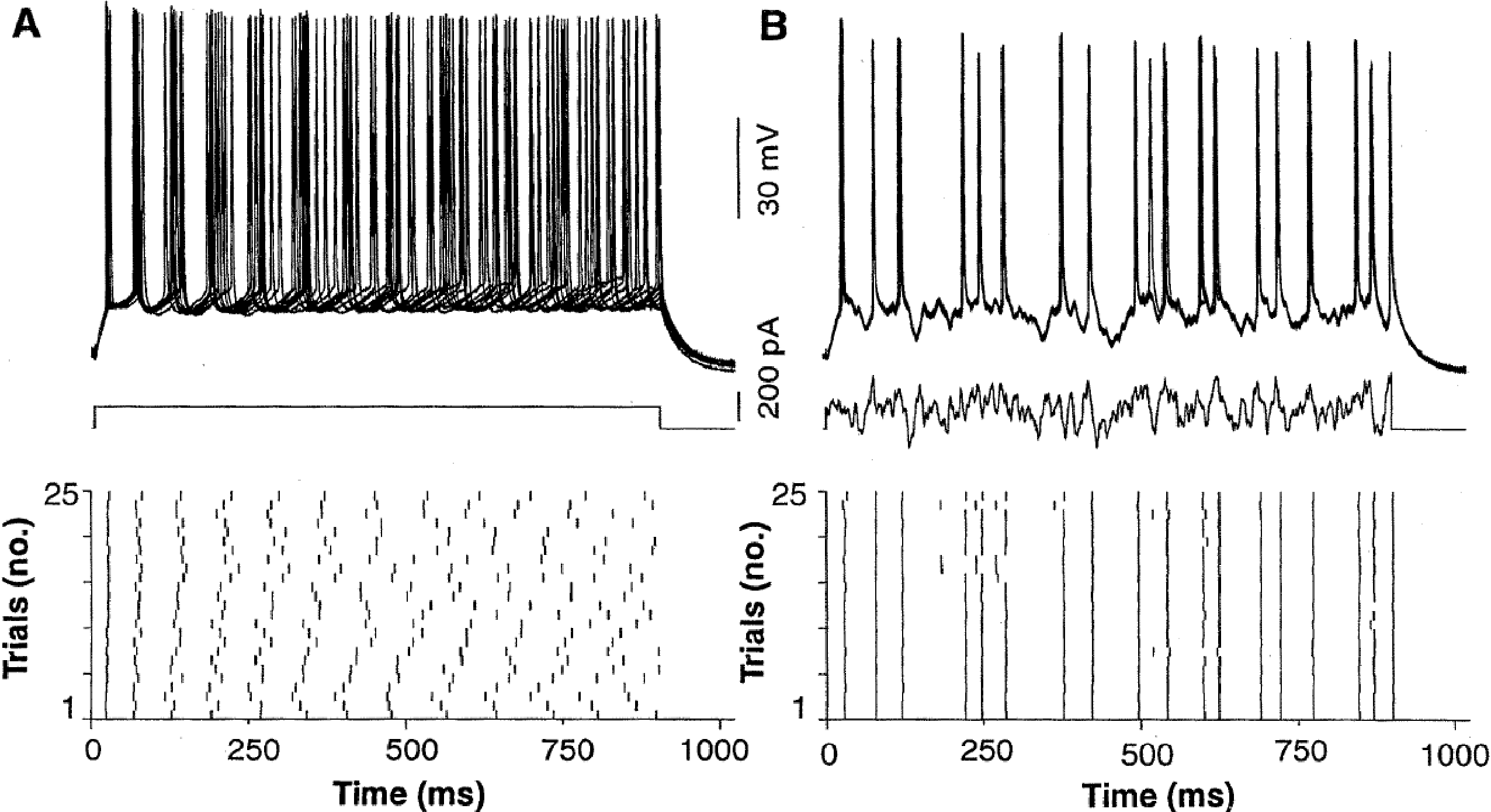}


\caption{
Reliability of spike timing in a rat neocortical neuron stimulated by (A) a constant supra-threshold current pulse; (B) a time-varying current (frozen filtered white Gaussian noise added to a constant component). 
Raster plots of the spike times for \num{25} consecutive trials are depicted at the bottom, for both experiments.
Reprinted from~\cite{Mainen1995}
with permission from the 
AAAS.
}
\label{fig_rats}
\end{figure}

Mainen and Sejnowski~\cite{Mainen1995}
recorded the responses of a single rat neocortical neuron subject to an excitation current injected \textit{in vitro}, as reproduced in \cref{fig_rats}.
\textcolor{h}{They reproduced Bryant and Segundo~\cite{Bryant1976} pioneer experiment, who studied the responses of \textit{Aplysia} (marine gastropod mollusk) neurons.} We distinguish between two configurations corresponding to two different stimuli: (A) a constant current pulse of relatively large amplitude; (B) a time-varying current, which is the sum of a constant component and \textcolor{h}{a} white Gaussian noise (low-pass filtered so as to faithfully emulate the action of the synapse preceding the neuron).
These experiments have been widely replicated since then and are often invoked in the discussions comparing spike rate-encoding theory to event-based information theory for biological brains~\cite{Brette2015}.

The neuron responses were recorded for \num{25} 
different
trials. In experiment (B), the input white Gaussian noise is said to be frozen,
i.e.  
exactly the \emph{same} realisation (signal trace) is injected for all the trials.
Of course, the intrinsic noise of the biological neuron is unpredictable, and is random from one trial to another.
From an experimental point of view, this intrinsic noise manifests itself in the form of \emph{spike time uncertainty}, which can be easily observed in the raster plot of the spike times (see bottom plot of \cref{fig_rats}(A)).
One goal of these experiments is to gain insight on the reliability of the output spike times, which encode the information, i.e. the trial-to-trial reproducibility against the noise inherent to the system.
The two experiments, which differ in the type of 
stimulus, yield radically different results 
that reveal two different dynamical regimes 
for
the neuron.

In (A), the biological neuron, excited by a constant current pulse of relatively large amplitude, operates above threshold and exhibits a typical self-running (i.e. unregulated) \emph{oscillatory behaviour}: the spike time uncertainty \emph{accumulates} over time.
This uncertainty is conventionally referred to as 
\emph{jitter} or \emph{phase noise} in the electronics literature~\cite{Hajimiri2002,DaDalt2018}.
The oscillator phase error can be modelled as a diffusion process due to the random noise~\cite{Bonnin2013,Bonnin2017},
which is inherent to any dissipative system~\cite{PRL2025}, whether biological (as is the case here), electronic, mechanical or 
of any other kind.

In (B), under time-\textcolor{l}{varying} 
stimulus, the spike times are almost perfectly aligned vertically for the \num{25} repeated trials, 
except for a few outliers. 
Unlike experiment (A), there is no phase uncertainty or jitter \textcolor{h}{accumulation} over time. This amazing observation suggests substantially different spiking dynamics. In (B), each spike may be regarded as a discrete \emph{event} presumably resulting from the transition from a stable steady state (in which the neuron operates below threshold and is silent) to a limit cycle, triggered by 
a temporary positive surge in the
input current,
before returning to its quiescent state.
In the absence of 
temporary
positive contribution of frozen white Gaussian noise to the current, the neuron naturally remains in its silent state.
Such mechanism
is the signature of an \emph{excitable behaviour}~\cite{Izhikevich2007_book}.
Fluctuating or sparse input signals
play a resetting 
role and significantly reduce the spike timing uncertainty.
This experiment highlights the promise in terms of reliability of the biologically-inspired event-based neuromorphic computing paradigm~\cite{Sepulchre2022_PIEEE}.

\section{Spike Timing in 
Analog Spiking Neurons: Numerical Investigation}
\label{section:Spike Timing in Analog Spiking Neurons: Numerical Investigation}


The central theme of this paper is the reproduction of the neuroscience experiments for an \emph{artificial} analog neuron, such as the CMOS implementation of \cref{fig_neuron}. 
\textcolor{l}{In this work, we use the same deterministic model and parameters as~\cite{Sourikopoulos2017} but extend it to incorporate the intrinsic noise physically generated within the transistors.}
\textcolor{l}{We rely on computation-intensive time-domain simulations, also referred to as transient noise simulations (\texttt{NOISETRAN} in SPICE) that are compatible with industrial transistor models.~\cite{SSE2023,TCAS2025}.}

This framework avoids the need to resort to simplified analytical models \textcolor{l}{and home-made implementations of numerical stochastic simulations} that would result in inaccurate and implausible predictions greatly differing from 
experimental data on silicon.

One challenge is to simulate the intrinsic transistor noise in a highly nonlinear circuit.
Whilst a detailed simulation methodology was described in dedicated electronics engineering publications~\cite{SSE2023,TCAS2025}, we summarise the main ingredient\textcolor{h}{s} here.
\textcolor{l}{The \texttt{NOISETRAN} tool proceeds schematically as follows:}
(i) add
an individual physical noise source in parallel with each MOS transistor \textcolor{l}{(whose state-dependent intensity is given by the foundry model)};
(ii) generate random values at each time step for the noise realisations;
(iii) perform a conventional time-domain simulation, which essentially consists of numerically solving the nonlinear differential-algebraic equations of the circuit, 
involving
the transistor dynamical model \textcolor{l}{provided by the foundry}. 
This digital twin 
framework
was exploited and validated in the past to study noise-induced stochastic transitions in bistable circuits~\cite{LCSS2024}.
A previous work~\cite{Querlioz2013} used similar tools to study stochastic resonance in analog spiking circuits.

{\color{l}
An insightful estimate of the intrinsic noise magnitude can be obtained from the standard deviation of the integrated thermal noise on some node voltage at equilibrium, $\sqrt{\kB T/C}$~\cite{SSE2023}, where $\kB$ is Boltzmann's constant, $T = \SI{300}{\kelvin}$ is the temperature and $C$ is the estimated node capacitance encompassing the intrinsic transistor capacitances. 
Even though the spiking neuron operates out of equilibrium over a whole cycle, this still provides an order of magnitude.
The effective membrane capacitance extracted in~\cite{LASCAS2026} then gives a standard deviation of about $\SI{0.7}{\milli\volt}$ while $\vM$ varies from $0$ to almost $\SI{200}{\milli\volt}$ in spiking operation. The intrinsic noise thus results in typical fluctuations just below $\SI{1}{\percent}$.
}

\subsection{Neuron in the Oscillatory Regime}

\begin{figure}[]
\captionsetup[subfigure]{singlelinecheck=off,justification=centering}
\newcommand\myfontsize{\small}
\newcommand\mytickfontsize{\scriptsize}
\captionsetup[subfigure]{skip=0pt}

\vspace{+2mm}

\begin{subfigure}[t]{\linewidth}
\renewcommand\myfontsize{\small}
\renewcommand\mytickfontsize{\scriptsize}
\myfontsize
\centering
\psfragscanon
\psfrag{vGK(t) [mV]}[cc][cc]{$\vGK(t) \, [\si{\milli\volt}]$}
\psfrag{0}[cc][cc]{\mytickfontsize$0$}
\psfrag{100}[cc][cc]{\mytickfontsize$100$}
\psfrag{200}[cc][cc]{\mytickfontsize$200$}
\psfrag{//}[cc][cc]{\small //}
\includegraphics[scale=1]{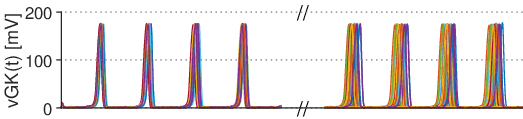}

\vspace{-2cm} 
\vspace{-\baselineskip} 
\vspace{+1mm} 
\subcaption{Output spikes}
\label{fig_MTNS2026_oscillating_vGK}

\end{subfigure}

\vspace{-4mm}

\begin{subfigure}[t]{\linewidth}
\renewcommand\myfontsize{\small}
\renewcommand\mytickfontsize{\scriptsize}
\myfontsize
\centering
\subcaption{Input excitation current $\iex(t)$}
\label{fig_MTNS2026_oscillating_iex}
\psfragscanon
\psfrag{0}[cr][cr]{\mytickfontsize$0$}
\psfrag{20 pA}[cr][cr]{\color{l}\mytickfontsize$20\,[\si{\pico\ampere}]$}
\includegraphics[scale=1]{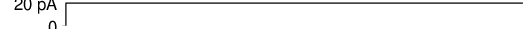}
\end{subfigure}

\vspace{+1mm}

\begin{subfigure}[t]{\linewidth}
\renewcommand\myfontsize{\small}
\renewcommand\mytickfontsize{\scriptsize}
\myfontsize
\centering
\psfragscanon
\psfrag{t [ms]}[cc][cc]{$t \, [\si{\milli\second}]$}
\psfrag{Trials}[cc][cc]{Trials}
\psfrag{1}[cc][cc]{\mytickfontsize$1$}
\psfrag{25}[cc][cc]{\mytickfontsize$25$}
\psfrag{0.00}[cc][cc]{\mytickfontsize$0$}
\psfrag{0.05}[cc][cc]{\mytickfontsize$0.05$}
\psfrag{0.10}[cc][cc]{\mytickfontsize$0.10$}
\psfrag{0.15}[cc][cc]{\mytickfontsize$0.15$}
\psfrag{0.85}[cc][cc]{\mytickfontsize$0.85$}
\psfrag{0.90}[cc][cc]{\mytickfontsize$0.90$}
\psfrag{0.95}[cc][cc]{\mytickfontsize$0.95$}
\psfrag{//}[cc][cc]{\small //}
\includegraphics[scale=1]{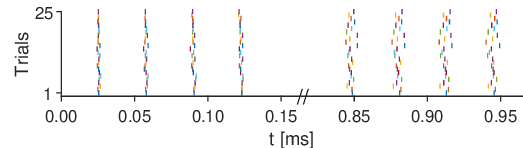}

\vspace{-2.5cm} 
\vspace{-\baselineskip} 
\vspace{+1mm} 
\subcaption{Spike times}
\label{fig_MTNS2026_oscillating_raster}

\end{subfigure}


\caption{
CMOS analog neuron in oscillatory regime under a constant excitation current pulse of supra-threshold amplitude $\SI{20}{\pico\ampere}$.
\subref{fig_MTNS2026_oscillating_vGK}
Output spikes.
\subref{fig_MTNS2026_oscillating_iex} Input excitation current $\iex(t)$.
\subref{fig_MTNS2026_oscillating_raster}
Raster plot of the spike times.
\newline
Simulation parameters: time step $\Delta t = \SI{1}{\nano\second}$, 
intrinsic transistor
noise bandwidth $\fmax = \SI{500}{\mega\hertz}$. \num{1000} trials were performed; only \num{25} are shown.
\vspace{-2mm}
}
\label{fig_MTNS2026_oscillating}
\end{figure}

In a recent work~\cite{LASCAS2026}, we numerically investigated the excitability property of the spiking neuron of \cref{fig_neuron} from an input-output characterisation, in the absence of intrinsic noise. 
Cold initial conditions are assumed, specifically
 $(\vM(0),\vGNA(0),\vGK(0)) = (\SI{0}{\volt},\SI{0}{\volt},\SI{0}{\volt})$
i.e. there is no initial residual charge in the circuit of \cref{fig_neuron}.
Inspired by~\cite{Sepulchre2018,Sepulchre2025}, we swept the amplitude and duration parameters of an input current pulse and measured for each case the output neuronal response, which can be reduced to the Boolean discrete event ‘spike or not?’. In particular, the obtained map (\cite[Figure~2(a)] {LASCAS2026}) tells us which pulse amplitude/duration combinations can lead to a spike. For the 
design of~\cite{Sourikopoulos2017} 

We depict in \cref{fig_MTNS2026_oscillating} the results of transient noise simulations of the neuron excited \emph{above} threshold by a constant current $\iex(t) = \SI{20}{\pico\ampere}$ (\cref{fig_MTNS2026_oscillating_iex}).
Following~\cite{Mainen1995}'s experiments, \num{25} repeated numerical trials were performed.
This sustained input current yields an action potential train, i.e. an output spike train on node $\vGK$ (all trials are superimposed in \cref{fig_MTNS2026_oscillating_vGK}). The mean spike period or inter-spike interval is found to be about $\SI{32}{\micro\second}$. Such high spike rate compared to biological neurons operating in the $\si{\milli\second}$ range is 
due
the small $RC$ time constant of CMOS analog neurons~\cite{Chicca2014_PIEEE}, especially for ULP implementations using small capacitances.

In the configuration of \cref{fig_MTNS2026_oscillating}, the neuron actually behaves like an oscillator, exactly like the rat neocortical neuron in \cref{fig_rats}(A). We can indeed observe an
increase in
spike time uncertainty as the variability of successive spike edges in \cref{fig_MTNS2026_oscillating_vGK}.
To quantify this effect, we have defined the successive spike times $T_1$, $T_2$,..., $T_n$ ($n$ reaching \num{30} for the $\SI{1}{\milli\second}$ simulation) as the 
times at which $\vGK$ crosses half of the supply voltage $\VDD/2 = \SI{100}{\milli\volt}$ on its rising edge.
We have extracted them for each trial, as shown in the raster plot in \cref{fig_MTNS2026_oscillating_raster}. The time axis was broken to clearly highlight that the spike time variability
or jitter
increases with the spike number.
All in all, \cref{fig_MTNS2026_oscillating} is precisely the counterpart of the neuroscience experiment (A) for the CMOS neuron in oscillatory regime. The jitter accumulation will be quantified and analysed in \cref{section:Results and Discussion}.

\subsection{Neuron Excited by a Time-Varying Input}

\begin{figure}[]
\captionsetup[subfigure]{singlelinecheck=off,justification=centering}
\newcommand\myfontsize{\small}
\newcommand\mytickfontsize{\scriptsize}
\captionsetup[subfigure]{skip=0pt}

\vspace{+2mm}

\begin{subfigure}[t]{\linewidth}
\renewcommand\myfontsize{\small}
\renewcommand\mytickfontsize{\scriptsize}
\myfontsize
\centering
\psfragscanon
\psfrag{vGK(t) [mV]}[cc][cc]{$\vGK(t) \, [\si{\milli\volt}]$}
\psfrag{0}[cc][cc]{\mytickfontsize$0$}
\psfrag{100}[cc][cc]{\mytickfontsize$100$}
\psfrag{200}[cc][cc]{\mytickfontsize$200$}
\makebox[\linewidth][l]{%
    \hspace*{-2.5mm}%
    \includegraphics[scale=1]{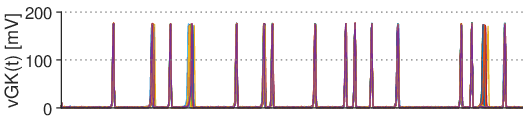}%
    }

\vspace{-2cm} 
\vspace{-\baselineskip} 
\vspace{+1mm} 
\subcaption{Output spikes}
\label{fig_MTNS2026_WGN_filtered_vGK}

\end{subfigure}

\vspace{+2mm}

\begin{subfigure}[t]{\linewidth}
\renewcommand\myfontsize{\small}
\renewcommand\mytickfontsize{\scriptsize}
\myfontsize
\centering
\psfragscanon
\psfrag{[pA]}[cr][cr]{\mytickfontsize$[\si{\pico\ampere}]$}
\psfrag{20}[cr][cr]{\mytickfontsize$20$}
\psfrag{0}[cr][cr]{\mytickfontsize$0$}
\psfrag{-20}[cr][cr]{\mytickfontsize$-20$}
\makebox[\linewidth][l]{%
    \hspace*{-2.5mm}%
    \includegraphics[scale=1]{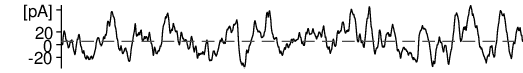}%
}

\vspace{-1.5cm} 
\vspace{-\baselineskip} 
\vspace{+4mm} 
\subcaption{Input excitation current $\iex(t)$}
\label{fig_MTNS2026_WGN_filtered_iex}

\end{subfigure}

\vspace{+2mm}

\begin{subfigure}[t]{\linewidth}
\renewcommand\myfontsize{\small}
\renewcommand\mytickfontsize{\scriptsize}
\myfontsize
\centering
\psfragscanon
\psfrag{t [ms]}[cc][cc]{$t \, [\si{\milli\second}]$}
\psfrag{Trials}[cc][cc]{Trials}
\psfrag{1}[cc][cc]{\mytickfontsize$1$}
\psfrag{25}[cc][cc]{\mytickfontsize$25$}
\psfrag{0}[cc][cc]{\mytickfontsize$0$}
\psfrag{0.2}[cc][cc]{\mytickfontsize$0.2$}
\psfrag{0.4}[cc][cc]{\mytickfontsize$0.4$}
\psfrag{0.6}[cc][cc]{\mytickfontsize$0.6$}
\psfrag{0.8}[cc][cc]{\mytickfontsize$0.8$}

\makebox[\linewidth][l]{%
    \hspace*{-2.5mm}%
    \includegraphics[scale=1]{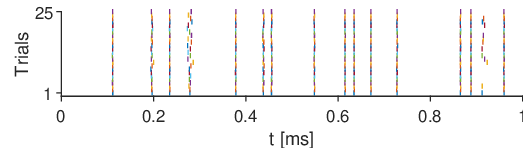}%
}

\vspace{-2.5cm} 
\vspace{-\baselineskip} 
\vspace{+1mm} 
\subcaption{Spike times}
\label{fig_MTNS2026_WGN_filtered_raster}

\end{subfigure}


\caption{
CMOS analog neuron excited by frozen filtered white Gaussian noise, fluctuation around a constant current pulse of sub-threshold amplitude $\SI{5}{\pico\ampere}$. Input noise parameters: standard deviation $\sigma_{\iex}  = \SI{20}{\pico\ampere}$, filter time constant $\tau = \SI{5}{\micro\second}$.
\subref{fig_MTNS2026_WGN_filtered_vGK}
Output spikes.
\subref{fig_MTNS2026_WGN_filtered_iex} Input excitation current $\iex(t)$.
\subref{fig_MTNS2026_WGN_filtered_raster}
Raster plot of the spike times.
Same simulation parameters as in \cref{fig_MTNS2026_oscillating}.
\vspace{-2mm}
}
\label{fig_MTNS2026_WGN_filtered}
\end{figure}

We are now interested in the configuration where the neuron is excited by a time-varying input current whose constant (DC) component is \emph{below} the excitability threshold of $\SI{7}{\pico\ampere}$.
In an attempt to replicate~\cite{Mainen1995}'s experiment (B), we have generated and injected a frozen filtered white Gaussian noise signal. The constant component of $\SI{5}{\pico\ampere}$ 
is on its own, as desired, insufficient to trigger the spiking mechanism that requires the contribution of the input fluctuations as well as, possibly, of the intrinsic noise of the circuit.
To emulate the low-pass filtering effect of the synapse that would provide the excitation current, 
an ideal (yet band-limited by the time step \textcolor{l}{given in the caption of \cref{fig_MTNS2026_oscillating}}) white Gaussian noise signal was convoluted with the impulse response indicated by~\cite{Mainen1995}:
\begin{equation}
\label{eq:h(t)}
h(t) = \frac{1}{\tau^2} \, t \, \exp\Big(-\frac{t}{\tau}\Big)
\text{,}
\end{equation}
in which the filter time constant $\tau = \SI{5}{\micro\second}$ was chosen relatively to the time scale of the CMOS neuron to obtain a signal pattern similar to~\cite{Mainen1995}.
The standard deviation of $\iex(t)$ is $\sigma_{\iex} = \SI{20}{\pico\ampere}$.
We further emphasise that the input $\iex(t)$ (\cref{fig_MTNS2026_WGN_filtered_iex}) is said to be ‘frozen’ in the sense that exactly the \emph{same} realisation is injected into the circuit of \cref{fig_neuron} for all 
simulation trials.
\textcolor{l}{Although it was built with a random procedure, the input can therefore be seen as a deterministic time-varying signal and is identical from one trial to another.}

We depict in \cref{fig_MTNS2026_WGN_filtered} the results of transient noise simulations.
Like in~\cite{Mainen1995}'s experiment of \cref{fig_rats}(B), the firing pattern of \cref{fig_MTNS2026_WGN_filtered_vGK} is irregular since an output spike is only produced when the input current fluctuations are sufficient to 
induce
an above-threshold response~\cite{Sepulchre2022_PIEEE}.
The neuron operates in the \emph{excitability regime} and does \emph{not} behave like a self-running oscillator any more.
As a consequence, the spike timing does not suffer from accumulated jitter as can be noticed in \cref{fig_MTNS2026_WGN_filtered_raster}: almost all the spike times are closely aligned vertically along the \num{25} trials.
Overall, \cref{fig_MTNS2026_WGN_filtered} is precisely the counterpart of the neuroscience experiment (B) for the CMOS neuron in the excitability regime.

We notice that the 4th and 15th spikes are naturally more prone to timing variability.
This observation suggests that the transition of the neuron from resting to spiking state is significantly 
influenced
by the intrinsic noise of the circuit, akin to the stochastic state transitions in bistable memories~\cite{LCSS2024,TCAS2025}. 
\textcolor{l}{The 4th and 15th spikes are the ones affected for the particular time-varying input of \cref{fig_MTNS2026_WGN_filtered_iex}.
To a smaller extent, the 2nd and 16th spikes also show greater variability.
}
Because the study of these special mechanisms falls beyond the scope of this paper, the 4th and 15th spikes will be regarded as outliers in the discussion of \cref{section:Results and Discussion}.

\section{Results and Discussion}
\label{section:Results and Discussion}

The numerical simulations of \cref{section:Spike Timing in Analog Spiking Neurons: Numerical Investigation} reproduced  the neuroscience experiments by~\cite{Mainen1995} on
a CMOS analog spiking neuron.
In this section, we quantitatively confirm the conclusions on the reliability of spike timing by empirically estimating the variance of the successive spike times.
\textcolor{h}{
The observed significant spike time variations in the oscillatory regime are \textcolor{l}{highlighted} as a physical limitation by a thermodynamic uncertainty relation.
}

{\color{h}
\subsection{Jitter Analysis}
}

\begin{figure}[]
\newcommand\myfontsize{\small}
\newcommand\mytickfontsize{\footnotesize}
\myfontsize
\centering
\psfragscanon
\psfrag{Nth spike}[cc][cc]{\myfontsize $n$th spike}
\psfrag{Var[TN-T1] [(us)^2]}[cc][cc]{\myfontsize$\Var\{T_n - T_1\} \, [(\si{\micro\second})^2]$}
\psfrag{0}[cc][cc]{\mytickfontsize$0$}
\psfrag{1}[cc][cc]{\mytickfontsize$1$}
\psfrag{2}[cc][cc]{\mytickfontsize$2$}
\psfrag{3}[cc][cc]{\mytickfontsize$3$}
\psfrag{4}[cc][cc]{\mytickfontsize$4$}
\psfrag{5}[cc][cc]{\mytickfontsize$5$}
\psfrag{6}[cc][cc]{\mytickfontsize$6$}
\psfrag{7}[cc][cc]{\mytickfontsize$7$}
\psfrag{8}[cc][cc]{\mytickfontsize$8$}
\psfrag{5}[cc][cc]{\mytickfontsize$5$}
\psfrag{10}[cc][cc]{\mytickfontsize$10$}
\psfrag{15}[cc][cc]{\mytickfontsize$15$}
\psfrag{20}[cc][cc]{\mytickfontsize$20$}
\psfrag{25}[cc][cc]{\mytickfontsize$25$}
\psfrag{30}[cc][cc]{\mytickfontsize$30$}
\psfrag{Oscillating regime}[tl][tl]{\myfontsize\color{r}
\setlength{\tabcolsep}{0pt}
\begin{tabular}{l}
Oscillatory regime\\ 
\cref{fig_MTNS2026_oscillating}
\end{tabular}
}
\psfrag{Accumulation}[tl][tl]{\myfontsize\color{darkOrange}
\setlength{\tabcolsep}{0pt}
\begin{tabular}{l}
Accumulated jitter\\ 
$(n-1)\cdot \Var\{T_2 - T_1\}$
\end{tabular}
}
\psfrag{Excitable regime}[cl][cl]{\myfontsize\color{b}
\setlength{\tabcolsep}{0pt}
\begin{tabular}{l}
Excitability regime\\ 
\cref{fig_MTNS2026_WGN_filtered}
\end{tabular}
}
\includegraphics[scale=1]{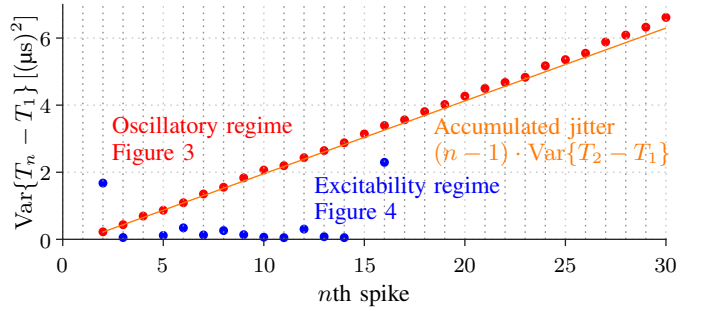}


\caption{
Analysis of the trial-to-trial variability in the spike times for the two types of input current, respectively triggering the oscillatory (\cref{fig_MTNS2026_oscillating}\textcolor{h}{, with mean inter-spike interval or spike period $\E\{T_n - T_{n-1}\} \approx \SI{32}{\micro\second}$}) and excitability (\cref{fig_MTNS2026_WGN_filtered}) behaviours. The jitter only accumulates over time 
for the former.
The variance was estimated from \num{1000} trials.
As discussed in the text, 
the variances of the 4th and 15th spike times of \cref{fig_MTNS2026_WGN_filtered}
are very large ($20$ and $8\,(\si{\micro\second})^2$) and go off the scale of the plot.
}
\label{fig_MTNS2026_jitter}
\end{figure}

In order to break free from the dependence on initial conditions and on the power-up phase of the circuit described in \cite{LASCAS2026}, we study the variability in the spike time \emph{differences} from the first spike (denoted $T_1$), that is we study $T_2 - T_1$, $T_3 - T_1$, ..., $T_n - T_1$, ... and we then compute $\Var\{T_n - T_1\}$ for $n \geq 2$ over all the trials.
This quantity is conventionally referred to as \emph{$n$-period (spike time) jitter} or \emph{accumulated jitter} in electronics engineering~\cite{DaDalt2018}.
To ensure a certain statistical accuracy for the empirical estimator, the variance was estimated from \num{1000} successive trials (although only \num{25} were shown in \cref{fig_MTNS2026_oscillating,fig_MTNS2026_WGN_filtered}).
The results are summarised in \cref{fig_MTNS2026_jitter}.

As observed in \cref{fig_MTNS2026_oscillating}, the jitter accumulates over the time in the oscillatory regime. Moreover, we have verified that, for this specific regime,
\begin{equation}
\label{eq:accumulation}
\Var\{T_n - T_1\} \approx (n-1)\cdot \Var\{T_2 - T_1\}
\ \text{for } n \geq 2
\end{equation}
where $\Var\{T_2 - T_1\} \approx \Var\{T_n - T_{n-1}\}$
is nothing but the \emph{period jitter}, that is the variance of the inter-spike interval.
With reference to \cref{fig_MTNS2026_jitter}, the very close agreement of \eqref{eq:accumulation} \textcolor{l}{(represented by the orange straight line)} confirms that the spike time jitter is dominated by \emph{white} (time-decorrelated) noise sources, corresponding to the thermal noise of the MOS transistors of \cref{fig_neuron}.
This could have been anticipated from the knowledge of the circuit bandwidth~\cite{TCAS2025}.
The slight deviation of the simulation results from \eqref{eq:accumulation} from $n \approx 25$ is likely the signature of time-correlated $1/f$ noise~\cite{Hajimiri2002}.
While the noise was 
here 
generated and simulated in nominal conditions, the large transistor-to-transistor variability in the low-frequency noise magnitude~\cite{Wirth2025}
does
nevertheless lead us to question the universality of \eqref{eq:accumulation}.


{\color{h}

\subsection{Thermodynamic Uncertainty Relation}

\textcolor{l}{
We now introduce a classic result of stochastic thermodynamic theory that we will apply to the jitter analysis to highlight the physical reliability-dissipation tradeoff.} Stochastic thermodynamics aims to extend the laws of conventional thermodynamics to situations in which random fluctuations (due to the intrinsic noise) are not negligible~\cite{Peliti2021,van2013stochastic}.
This theory is especially capable of describing nonlinear systems operating out of equilibrium~\cite{Rao2016}.
Important instances include nanoscale CMOS circuits, either digital gates or memories~\cite{Freitas2021stochastic,EDTM2024,APL2023} or analog circuits exhibiting substantial nonlinearities like the spiking neuron of \cref{fig_neuron}.

A key result is the \emph{local detailed balance} relation~\cite{maes2021local,van2013stochastic}, which we now state in the specific formulation tailored to our needs.
Let a physical system be described by an overdamped time-homogeneous Markov process (in continuous time and discrete or continuous space) in 
interaction
with a thermal bath 
at
temperature $T$. Overdamped here means that the state space encodes even observables of the physical systems, i.e. observables that do not change sign under time reversal --- like position or voltage, and unlike speed or electric current that are odd observables changing sign under time reversal. An $RC$ circuit, linear or not, is an example of overdamped system, as are biological neurons. 
In particular, the dynamics of the neuronal circuit of \cref{fig_neuron} is overdamped since it is only made of nonlinear dissipative components and capacitances (not only $C_{\mathrm{m}}$ and $C_{\mathrm{K}}$ but also the internal capacitances of the transistors modelling their dynamical currents) without including inductors.
Let $\omega$ be a trajectory of such a system over a time interval $[0,t]$. A trajectory consists of an initial state $x_0$, followed by a sequence of (discrete or continuous) time-stamped transitions to other states, until a state $x_t$. This includes discrete state spaces (described by continuous-time Markov chains) and continuous ones (described by stochastic differential equations). Let $\overline{\omega}$ be the time-reversed path, i.e. the reverse sequence of states and transitions between them, starting from $x_t$ until $x_0$. Let $p(\omega)$ be the conditional probability (or probability density) of trajectory $\omega$ occurring knowing the initial state $x_0$. Likewise, $p(\overline{\omega})$ is the conditional probability (or probability density) of $\overline{\omega}$, knowing it starts at $x_t$. Let $E_{\text{dissip}}(\omega)$ be the energy dissipated into the heat bath (for instance, the energy dissipated into the resistive components of an $RC$ circuit) by trajectory $\omega$.
Then the local detailed balance relation writes \textcolor{l}{~\cite{maes2021local,van2013stochastic}}:
\begin{equation}
		\frac{p(\omega)}{p(\overline{\omega})}= 
		\exp\Big(-\frac{E_{\mathrm{dissip}} (\omega)}{\kB T}\Big)
\end{equation}
where $\kB$ is Boltzmann's constant. 
This principle applies broadly, in particular for nonlinear systems out of equilibrium. Similar formulations exist for more general contexts (e.g. in systems with several temperatures, underdamped systems, etc.). It can be seen as a generalisation of detailed balance principle (or reversibility for Markov processes), which occurs when the system is at thermodynamic equilibrium with the bath. It has been used to derive general results relating dynamical and thermodynamic quantities in nanoscale systems. 

One of them is the \emph{thermodynamic uncertainty relation} (TUR), which we now state. It applies to any odd observable $f(\omega)$ (i.e. a quantity, associated to the state transitions, that switches sign when the course of time is reversed) 
which is the result of a sum or an integral calculated across the trajectory.
 For instance $f$ can be the total charge (integral of current) going through a conductor of an overdamped circuit during the interval $[0,t]$, or the total energy dissipated (integral of power) in this circuit over the same time window, etc. Then the TUR states that for overdamped Markov systems with constant parameters at stationarity \cite{barato2015thermodynamic,Horowitz2020,falasco2020unifying}:
\begin{equation}
\label{eq:TUR1}
	\frac{\Var\{f\}}{\E\{f\}^2}\geq \frac{2\kB T}{\overline{E}_{\mathrm{dissip}}}
\end{equation}
where $\overline{E}_{\mathrm{dissip}} = \E\{E_{\mathrm{dissip}} (\omega)\}$ is the expected (i.e. averaged over all the trajectories $\omega$) energy dissipated into the bath over time interval $[0,t]$. 

This important result states that a high precision (low variance for a given mean) on $f$ can only be achieved at the cost of high dissipation $\overline{E}_{\mathrm{dissip}}$. 

\textcolor{l}{Let $N$ be the net number of spikes recorded in the circuit of \cref{fig_neuron} during the time interval $[0,t]$, corresponding to the net number of limit cycles followed by the trajectory in the state space of the model
(counting as a $-1$ for each negative spike if the direction of the trajectory is reversed).}
Let us now normalise it to $f=N/t$. This odd observable is effectively an empirical frequency for the circuit of \cref{fig_neuron}.
Thus the TUR \eqref{eq:TUR1} applied to $f=N/t$ indicates that an overdamped Markovian cyclic process, induced by constant external forces, which dissipates little energy cannot be precise,
thus exhibiting a large spike time uncertainty (i.e. jitter if regarded as a clock).

The expected energy dissipated per cycle in the case of the circuit of \cref{fig_neuron} is computed as 
\begin{equation}
	\label{eq:Edissip}
\overline{E}_{\mathrm{dissip-per-cycle}}
	= \int_{\mathrlap{\text{one spike cycle}}} 
	\VDD \cdot \E\{\iDD(t)\} \, \dt
		\text{,}
\end{equation}
 i.e. by integrating the total instantaneous supplied power (current) over a full spike cycle. \textcolor{l}{We have found} it to be $\approx \SI{1}{\femto\joule\per spike}$ when the neuron of \cref{fig_neuron} is excited by a constant current of $\SI{20}{\pico\ampere}$ in \cref{fig_MTNS2026_oscillating}.
The energy dissipated into the bath over time interval $[0,t]$ is on average, for large $t$:
\begin{equation}
	\label{eq:Edissiptot}
	\overline{E}_{\mathrm{dissip}} = \E \{N\} \cdot \overline{E}_{\mathrm{dissip-per-cycle}}
	\text{.}
\end{equation}
This therefore implies, for large $t$ (in particular for $t \to \infty$) a lower bound on the variance of the empirical frequency:
\begin{equation}
	\label{eq:TUR2}
	\frac{\Var\{N/t\}}{(\E \{N/t\})^2}\geq \frac{\SI{2.6e-10}{\second}}{t}
	\text{.}
\end{equation}

The jitter is captured by another quantity in \cref{fig_MTNS2026_jitter}, namely by the variance of $(T_n-T_1)$. Even though this is not mathematically equivalent to the variance of $f=N/t$ as appears in \eqref{eq:TUR2}, one may still take one as a proxy for the other and compare the respective orders of magnitude. From the data shown in \cref{fig_MTNS2026_oscillating} we find
{
\begin{equation}
	\label{eq:TUR3}
\frac{\Var\{T_n - T_{1}\}}{(\E\{T_n - T_{1}\})^2} \approx \frac{\num{2e-4}}{n-1}
\text{.}
\end{equation}
}
The right hand side of \eqref{eq:TUR3} is  a factor $\num{25}$ above the fundamental lower bound \eqref{eq:TUR1} (up to the approximations being made and taking $t = (n-1) \cdot \SI{32}{\micro\second}$).

The key point is that, irrespective of the particular choice of the circuit parameters for the simulation of \cref{fig_MTNS2026_oscillating} (e.g. the current amplitude), the TUR places a fundamental physical limit in terms of reliability-dissipation tradeoff: lowering the minimum achievable jitter necessarily requires a higher dissipation.

}


\subsection{Discussion}

In the excitability regime, the jitter remains very low for most of the spikes, and substantially below the jitter of the corresponding spikes when the neuron constantly oscillates.
Whilst a detailed study of the neuronal dynamics will be the object of a future work, we presumably attribute this interesting result to the fundamentally different spiking mechanism described at the end of \cref{section:Spike Timing in Biological Neurons: Neuroscience Experiments}: instead of accumulating phase error during repeated 
cycles, each spike is a discrete event resulting from a state transition triggered by a time-varying input, whose resetting action drastically reduces the jitter.
%
This preliminary numerical study suggests that a possibly random
time-varying
input signal could be the solution to an optimal control problem whose objective would be to minimise jitter (as well as the energy consumption) in a reliability perspective.
We finally mention that the TUR \eqref{eq:TUR2} does not apply to this configuration where the spiking is a non-stationary process. In other words, the physical limitations of the oscillatory regime, as imposed by thermodynamics, are alleviated in the excitability regime.


\section{Conclusions and Perspectives}
\label{section:Conclusions and Perspectives}

Relying on an industrial circuit simulation framework, we were able to reproduce famous neuroscience experiments leveraging the excitability of a CMOS 
leaky integrate-and-fire
neuron.
The intrinsic transistor noise causes variability in the spike times across different trials, a phenomenon referred to as jitter.
When the neuron is excited by a constant supra-threshold current, it exhibits a typical self-running oscillatory behaviour with linear accumulation of jitter over time.
Such expected and observed behaviour is also supported by a thermodynamic uncertainty relation descending from stochastic thermodynamics that highlights the reliability-dissipation tradeoff in this regime.
By contrast, a time-varying input fundamentally modifies the neuron spiking mechanism, whereby each spike results from a controlled transition, which 
substantially reduces jitter.
These rigorous numerical experiments further pinpoint the potential of the biologically-inspired event-based neuromorphic computing paradigm given its assets in terms of noise immunity, in addition to the expected energy saving.
In a future work, we will model the neuronal dynamics to validate the assumptions raised on the different spiking mechanisms and to predict the 
underlying 
spike time statistics.


\section*{Acknowledgement}

First of all, the authors would like to thank Prof. Rodolphe Sepulchre for bringing the rat experiment from~\cite{Mainen1995}
to our attention and for motivating the study of the excitability of CMOS analog spiking neurons.
We also thank Prof. Fernando Corinto, Prof. 
Richard 
Stanley Williams, Prof. Pietro Maris Ferreira, Dr. Martin Lefebvre, Dr. Meik Dörpinghaus and Mr. Alexandre Thyrion for the fruitful discussions and for providing relevant references that 
allowed us to improve
this work and to open perspectives for future research.


\bibliographystyle{IEEEtran}
\bibliography{IEEEabrv,bib}

\begin{thebibliography}{10}
\providecommand{\url}[1]{#1}
\csname url@samestyle\endcsname
\providecommand{\newblock}{\relax}
\providecommand{\bibinfo}[2]{#2}
\providecommand{\BIBentrySTDinterwordspacing}{\spaceskip=0pt\relax}
\providecommand{\BIBentryALTinterwordstretchfactor}{4}
\providecommand{\BIBentryALTinterwordspacing}{\spaceskip=\fontdimen2\font plus
\BIBentryALTinterwordstretchfactor\fontdimen3\font minus
  \fontdimen4\font\relax}
\providecommand{\BIBforeignlanguage}[2]{{%
\expandafter\ifx\csname l@#1\endcsname\relax
\typeout{** WARNING: IEEEtran.bst: No hyphenation pattern has been}%
\typeout{** loaded for the language `#1'. Using the pattern for}%
\typeout{** the default language instead.}%
\else
\language=\csname l@#1\endcsname
\fi
#2}}
\providecommand{\BIBdecl}{\relax}
\BIBdecl

\bibitem{Sepulchre2022_PIEEE}
R.~Sepulchre, ``Spiking control systems,'' \emph{Proceedings of the IEEE}, vol.
  110, no.~5, pp. 577--589, 2022.

\bibitem{Chicca2014_PIEEE}
E.~Chicca, F.~Stefanini, C.~Bartolozzi, and G.~Indiveri, ``Neuromorphic
  electronic circuits for building autonomous cognitive systems,''
  \emph{Proceedings of the IEEE}, vol. 102, no.~9, pp. 1367--1388, 2014.

\bibitem{Querlioz2013}
D.~Querlioz and V.~Trauchessec, ``Stochastic resonance in an analog
  current-mode neuromorphic circuit,'' in \emph{2013 IEEE International
  Symposium on Circuits and Systems (ISCAS)}.\hskip 1em plus 0.5em minus
  0.4em\relax IEEE, 2013, pp. 1596--1599.

\bibitem{Sourikopoulos2017}
I.~Sourikopoulos, S.~Hedayat, C.~Loyez, F.~Danneville, V.~Hoel, E.~Mercier, and
  A.~Cappy, ``{A 4-fJ/spike artificial neuron in 65 nm CMOS technology},''
  \emph{Frontiers in neuroscience}, vol.~11, p. 123, 2017.

\bibitem{ascoli2025edge}
A.~Ascoli, A.~S. Demirkol, I.~Messaris, V.~Ntinas, D.~Prousalis, S.~Slesazeck,
  T.~Mikolajick, F.~Corinto, M.~Bonnin, M.~Gilli \emph{et~al.}, ``Edge of chaos
  theory unveils the first and simplest ever reported {H}odgkin--{H}uxley
  neuristor,'' \emph{Advanced Electronic Materials}, vol.~11, no.~8, p.
  2400789, 2025.

\bibitem{zhao2025}
R.~Zhao, T.~Wang, T.~Moon, Y.~Xu, J.~Zhao, P.~Sud, S.~J. Kim, H.-T. Liao,
  Y.~Zhuo, R.~Midya \emph{et~al.}, ``A spiking artificial neuron based on one
  diffusive memristor, one transistor and one resistor,'' \emph{Nature
  Electronics}, pp. 1--11, 2025.

\bibitem{Greatorex2025}
H.~Greatorex, O.~Richter, M.~Mastella, M.~Cotteret, P.~Klein, M.~Fabre,
  A.~Rubino, W.~Soares~Gir{\~a}o, J.~Chen, M.~Ziegler \emph{et~al.}, ``{A
  neuromorphic processor with on-chip learning for beyond-CMOS device
  integration},'' \emph{Nature communications}, vol.~16, no.~1, p. 6424, 2025.

\bibitem{LASCAS2026}
L.~Van~Brandt, G.~Brandsteert, and D.~Flandre, ``{On the Excitability of
  Ultra-Low-Power CMOS Analog Spiking Neurons},'' in \emph{IEEE LASCAS 2026,
  Arequipa, Peru}, 2026.

\bibitem{SSE2023}
L.~Van~Brandt, F.~Silveira, J.-C. Delvenne, and D.~Flandre, ``{On Noise-Induced
  Transient Bit Flips in Subthreshold SRAM},'' \emph{Solid-State Electronics},
  vol. 208, p. 108715, 2023.

\bibitem{TCAS2025}
L.~Van~Brandt, M.~Bonnin, M.~B. da~Silva, P.~Bolcato, G.~I. Wirth, D.~Flandre,
  and J.-C. Delvenne, ``{Modeling and Predicting Noise-Induced Failure Rates in
  Ultra-Low-Voltage SRAM Bitcells Affected by Process Variations},'' \emph{IEEE
  Transactions on Circuits and Systems I: Regular Papers}, 2025.

\bibitem{Bryant1976}
H.~L. Bryant and J.~P. Segundo, ``Spike initiation by transmembrane current: a
  white-noise analysis.'' \emph{The Journal of physiology}, vol. 260, no.~2,
  pp. 279--314, 1976.

\bibitem{Mainen1995}
Z.~F. Mainen and T.~J. Sejnowski, ``Reliability of spike timing in neocortical
  neurons,'' \emph{Science}, vol. 268, no. 5216, pp. 1503--1506, 1995.

\bibitem{Kirby2021}
T.~M. Kirby, L.~Ribar, and R.~Sepulchre, ``{Reliability of Event Timing in
  Silicon Neurons},'' \emph{arXiv preprint arXiv:2112.14134}, 2021.

\bibitem{Lansky1999}
P.~L{\'a}nsk{\`y} and R.~Rodriguez, ``Two-compartment stochastic model of a
  neuron,'' \emph{Physica D: Nonlinear Phenomena}, vol. 132, no. 1-2, pp.
  267--286, 1999.

\bibitem{Shimokawa1999}
T.~Shimokawa, K.~Pakdaman, and S.~Sato, ``Time-scale matching in the response
  of a leaky integrate-and-fire neuron model to periodic stimulus with additive
  noise,'' \emph{Physical Review E}, vol.~59, no.~3, p. 3427, 1999.

\bibitem{Bonnin2025_TCAS}
M.~Bonnin, L.~Van~Brandt, J.-C. Delvenne, F.~L. Traversa, and F.~Bonani, ``{A
  Thermodynamically Consistent Thermal Equilibrium Gaussian White Noise Model
  for Nonlinear Resistors},'' \emph{IEEE Transactions on Circuits and Systems
  I: Regular Papers}, 2025.

\bibitem{Brette2015}
R.~Brette, ``{Philosophy of the Spike: Rate-Based vs. Spike-Based Theories of
  the Brain},'' \emph{Frontiers in systems neuroscience}, vol.~9, p. 151, 2015.

\bibitem{Hajimiri2002}
A.~Hajimiri, S.~Limotyrakis, and T.~H. Lee, ``{Jitter and Phase Noise in Ring
  Oscillators},'' \emph{IEEE Journal of Solid-State Circuits}, vol.~34, no.~6,
  pp. 790--804, 2002.

\bibitem{DaDalt2018}
N.~Da~Dalt and A.~Sheikholeslami, \emph{Understanding jitter and phase noise: A
  circuits and systems perspective}.\hskip 1em plus 0.5em minus 0.4em\relax
  Cambridge University Press, 2018.

\bibitem{Bonnin2013}
M.~Bonnin and F.~Corinto, ``{Phase Noise and Noise Induced Frequency Shift in
  Stochastic Nonlinear Oscillators},'' \emph{IEEE Transactions on Circuits and
  Systems I: Regular Papers}, vol.~60, no.~8, pp. 2104--2115, 2013.

\bibitem{Bonnin2017}
M.~Bonnin, ``Amplitude and phase dynamics of noisy oscillators,''
  \emph{International Journal of Circuit Theory and Applications}, vol.~45,
  no.~5, pp. 636--659, 2017.

\bibitem{PRL2025}
J.-C. Delvenne and L.~Van~Brandt, ``{Moments of Entropy Production in
  Dissipative Devices},'' \emph{Physical Review Letters}, vol. 134, no.~23, p.
  237101, 2025.

\bibitem{Izhikevich2007_book}
E.~M. Izhikevich, \emph{{Dynamical Systems in Neuroscience}}.\hskip 1em plus
  0.5em minus 0.4em\relax MIT press, 2007.

\bibitem{LCSS2024}
L.~Van~Brandt and J.-C. Delvenne, ``{Predicting State Transitions in Autonomous
  Nonlinear Bistable Systems with Hidden Stochasticity},'' \emph{IEEE Control
  Systems Letters}, 2024.

\bibitem{Sepulchre2018}
R.~Sepulchre, G.~Drion, and A.~Franci, ``Excitable behaviors,'' in
  \emph{Emerging Applications of Control and Systems Theory: A Festschrift in
  Honor of Mathukumalli Vidyasagar}.\hskip 1em plus 0.5em minus 0.4em\relax
  Springer, 2018, pp. 269--280.

\bibitem{Sepulchre2025}
R.~Sepulchre and G.~Tong, ``On the threshold of excitable systems: An
  energy-based perspective,'' in \emph{2025 IEEE 64th Conference on Decision
  and Control (CDC)}, 2025, pp. 980--985.

\bibitem{Wirth2025}
G.~I. Wirth, T.~H. Both, M.~B. Silva, and L.~Van~Brandt, ``{Statistical
  Modeling of Timing Variability Due to Random Telegraph Noise in Logic
  Gates},'' \emph{Fluctuation and Noise Letters}, p. 2540022, 2025.

\bibitem{Peliti2021}
L.~Peliti and S.~Pigolotti, \emph{{Stochastic Thermodynamics: An
  Introduction}}.\hskip 1em plus 0.5em minus 0.4em\relax Princeton University
  Press, 2021.

\bibitem{van2013stochastic}
C.~Van~den Broeck \emph{et~al.}, ``Stochastic thermodynamics: A brief
  introduction,'' \emph{Phys. Complex Colloids}, vol. 184, pp. 155--193, 2013.

\bibitem{Rao2016}
R.~Rao and M.~Esposito, ``Nonequilibrium thermodynamics of chemical reaction
  networks: Wisdom from stochastic thermodynamics,'' \emph{Physical Review X},
  vol.~6, no.~4, p. 041064, 2016.

\bibitem{Freitas2021stochastic}
N.~Freitas, J.-C. Delvenne, and M.~Esposito, ``{Stochastic Thermodynamics of
  Nonlinear Electronic Circuits: A Realistic Framework for Computing Around
  kT},'' \emph{Physical Review X}, vol.~11, no.~3, p. 031064, 2021.

\bibitem{EDTM2024}
L.~Van~Brandt, D.~Flandre, and J.-C. Delvenne, ``{Stochastic Nonlinear
  Dynamical Modelling of SRAM Bitcells in Retention Mode},'' in \emph{2024 8th
  IEEE Electron Devices Technology \& Manufacturing Conference (EDTM),
  Bangalore, India}.\hskip 1em plus 0.5em minus 0.4em\relax IEEE, 2024, invited
  talk.

\bibitem{APL2023}
L.~Van~Brandt and J.-C. Delvenne, ``Noise-dissipation relation for nonlinear
  electronic circuits,'' \emph{Applied Physics Letters}, vol. 122, no.~26,
  2023.

\bibitem{maes2021local}
C.~Maes, ``Local detailed balance,'' \emph{SciPost Physics Lecture Notes}, p.
  032, 2021.

\bibitem{barato2015thermodynamic}
A.~C. Barato and U.~Seifert, ``Thermodynamic uncertainty relation for
  biomolecular processes,'' \emph{Physical review letters}, vol. 114, no.~15,
  p. 158101, 2015.

\bibitem{Horowitz2020}
J.~M. Horowitz and T.~R. Gingrich, ``Thermodynamic uncertainty relations
  constrain non-equilibrium fluctuations,'' \emph{Nature Physics}, vol.~16,
  no.~1, pp. 15--20, 2020.

\bibitem{falasco2020unifying}
G.~Falasco, M.~Esposito, and J.-C. Delvenne, ``Unifying thermodynamic
  uncertainty relations,'' \emph{New Journal of Physics}, vol.~22, no.~5, p.
  053046, 2020.

\end{thebibliography}


\end{document}